\documentclass[12pt,aps]{revtex4}
\usepackage{epsfig}
\usepackage{amsmath,amssymb,color}
\usepackage{epstopdf}
\parskip=\medskipamount

\newcommand{\be}{\begin{equation}}
\newcommand{\ee}{\end{equation}}

\newcommand{\fig}[1]{Fig.\ref{#1}}

\begin{document}

\title{Many-body localization and new critical phenomena in regular random graphs and constrained
Erd\H{o}s-R\'enyi networks}

\author{V. Avetisov$^{1,2}$, A. Gorsky$^{3,4}$, S. Nechaev$^{5,6}$, and O. Valba$^{1,2}$}

\affiliation{$^1$ N.N. Semenov Institute of Chemical Physics of the Russian Academy of Sciences,
119991, Moscow, Russia, \\
$^2$ Department of Applied Mathematics, National Research University Higher School of Economics,
101000, Moscow, Russia, \\
$^3$ Institute of Information Transmission Problems of the Russian Academy of Sciences, Moscow,
Russia, \\
$^4$ Moscow Institute of Physics and Technology, Dolgoprudny 141700, Russia \\
$^5$ Interdisciplinary Scientific Center Poncelet (CNRS UMI 2615), Moscow, Russia \\
$^6$ P.N. Lebedev Physical Institute of the Russian Academy of Sciences, 119991, Moscow, Russia}

\begin{abstract}

We consider from the localization perspective the new critical phenomena discovered recently for perturbed random regular graphs (RRG) and constrained Erd\H{o}s-R\'enyi networks (CERN) \cite{crit2}. At some critical value of the chemical potential of 3-cycles, $\mu$, the network decays into the maximally possible number of almost full subgraphs, and the spectrum of the Laplacian matrix acquires the two-zonal structure with a large gap. We find that the Laplacian eigenvalue statistics corresponds to delocalized states in one zone, and to the localized states in the second one. We interpret this behavior in terms of the many-body localization problem where the structure of the Fock space of some interacting many-body system is approximated by the RGG and/or by the CERN. We associate 3-cycles in RRGs and CERNs as resonant triples in the Fock space. We show that the scenario of the "localization without disorder", discussed previously in physical space, can be realized in the Fock space as well. We argue that it is natural to identify clusters in a RRG with particles in a many-body system above the phase transition. We discuss the controversial issue of an additional phase transition between ergodic and non-ergodic regimes in the delocalized phase in the Fock space and find a strong "memory dependence" of the states in the delocalized phase, thus advocating existence of non-ergodic delocalized states.

\end{abstract}

\maketitle

\section{Introduction}

Anderson one-particle localization \cite{anderson} provides the unified framework for the disorder-induced metal-insulator transitions in various physical problems. It was found that above some critical value of the diagonal disorder, the wave function becomes exponentially decaying in space and the system behaves as insulator. There are several criteria of the localization transition: the behavior of the "inverse participation ratio" of the wave function, the statistics of level spacing and the behavior of the entanglement entropy (see \cite{mirlinrev} for review). In this paper we are focused at the level spacing distribution. In the delocalized regime the distribution between nearest energy levels shares the Wigner-Dyson (surmise) law, while in the localized regime it follows the Poisson statistics. In $D>2$, the localized and delocalized states can coexist in different parts of the spectrum and are separated by the \emph{mobility edge}.

The concept of the one-body Anderson localization acquired recently a new incarnation in physics of interacting many-body systems \cite{basko,mirlin} (see \cite{huse,altman} for reviews of many-body localization). It implies that the effects of interaction strongly affect the transport properties of the statistical system, preventing it from the thermalization. The non-trivial trick had been suggested in \cite{kamenev}, where the many-body localization in the coordinate space was mapped onto the one-particle localization in the Fock space. According this hypothesis, the localization occurs in the Fock (Hilbert) space of a many-body system, approximated by a Bethe lattice, for which some exact results are available \cite{tau,efetov, ver,fyodorov}. Certainly, this mapping is quite approximate and schematically goes as follows \cite{kamenev}. The nodes of the Bethe lattice constitute the basis states in the Hilbert space, while the links correspond to the resonant pairs of states. If the wave function of an effective one-particle system in the Fock space is close to the state of the initial many-body system, then the localization in the Fock space occurs. Hence, one can identify the localized state in the Fock space with the particle in the initial many-body system. Meanwhile, if the wave function of the effective one-particle system in the Fock space is expanded over a large number of states of the many-body system, this regime is understood as a delocalized in the Fock space. Similarly to the localization in the real space, the notion of a mobility edge can be introduced in the Fock space as well.

Statistical properties of Bethe lattices have a lot in common with properties of random regular graphs (RRG), and the issue of the Fock space localization in terms of one-particle excitations on RRG attracts nowadays much attention \cite{krav,tikhonov,mirlin2,ioffe}. However it was not clear if the presence of cycles, which exist in RRGs and are absent in Bethe lattices, influence the one-particle localization on the network. There are controversial claims concerning this issue. Our study demonstrates that 3-cycles are of crucial importance for this issue and generically statistical models on RRG and on Bethe lattice can behave differently.

Recently, a surprising indication of an additional phase transition inside the delocalized regime was reported in \cite{biroli,krav,delu,biroli2}. It has been found in \cite{biroli,krav} that the critical value of the diagonal disorder separates the "ergodic" and "non-ergodic" delocalized regimes. In the non-ergodic case the absence of thermalization makes the very notion of equilibrium thermodynamics meaningless. This issue has been discussed in several papers, however no rigorous proof of existence or absence of the non-ergodic delocalized phase is yet provided. The non-ergodic delocalized phase has been related recently to the one-step replica symmetry breaking (RSB) in disordered system, and a kind of an analytic approach to this problem has been proposed in \cite{ioffe}. It was conjectured in \cite{ioffe} that the non-ergodic case corresponds to the RSB phase, while ergodic case means the unbroken replica symmetry. The multifractality exponents exhibit a jump at some critical value of the disorder signaling that the ergodic/non-ergodic transition happens inside the delocalized phase.

Another scenario of the localization was suggested long time ago \cite{kagan} and attracts much attention nowadays \cite{huv1,huv2}. It is usually called as "localization without disorder" or "localization in translational-invariant systems". The key point is the presence in the system the clusters of some nature. These clusters can be located everywhere in the space hence the translational invariance holds. No special on-site disorder is added to the system. It was shown in \cite{kagan} that clusters can provide the localization in the system and, hence, strongly influence the thermalization. Two possibilities have been discussed: i) either a single huge immobile cluster can be created, which influences thermalization; ii) or several mobile clusters can emerge whose motion affects thermalization (see \cite{huvrev} for review).

In the present work we exploit our recent findings concerning the critical behavior of CERN and RRG to gain new insight on the one-particle Anderson localization in the Fock space. Let us emphasize that the networks we consider in this study, belong to the so-called "mixed ensemble" where the number of links is fixed as in microcanonical ensemble, while the number of 3-cycles is controlled by the chemical potential as in the canonical ensemble. Critical properties of this ensemble differ from the ones of the canonical ensemble in which chemical potentials both for numbers of links and 3-cycles are introduced \cite{strauss}, as well as from the microcanonical ensemble in which both these numbers are fixed \cite{radin}.

We interpret the phase transition (found previously in \cite{crit2}) for CERN and RRG, controlled by the chemical potential for 3-cycles, in terms of the many-body localization in the physical space, and discuss related ergodic properties. The summary of main results obtained in \cite{crit2} is as follows:
\begin{itemize}
\item Above some critical value of the chemical potential, $\mu$, controlling the number of 3-cycles in the network, both CERN, and RRG get defragmented into the loosely connected collection of almost full subgraphs ("cliques");
\item The spectral densities of CERN and RRG form two-zonal structure. In particular:
\begin{itemize}
\item[--] The spectral density in the main "perturbative" zone above the transition point acquires triangle-like shape both for CERN and RRG. This spectral density differs from the one at $\mu=0$;
\item[--] The second "non-perturbative" zone in the spectral density, emerging for $\mu>\mu_{cr}$, is filled by eigenvalues corresponding to cliques (one eigenvalue per one clique).
\end{itemize}
\end{itemize}

In this paper we analyze numerically two questions:
\begin{itemize}
\item Which is the spectral statistics of RRG and CERN in the main and non-perturbative zones?
\item Do we have any signature of an intermediate non-ergodic behavior in the delocalized phase of our networks?
\end{itemize}

We demonstrate that statistics in the main zone corresponds to delocalized states while in the second (non-perturbative) zone -- to localized ones. We have no diagonal disorder in RRG and CERN, still the network gets completely defragmentated into clusters above $\mu_{cr}$. Just these clusters are responsible for the localization phenomena and we obtain the "localization without disorder" scenario in the Fock space which is the analog of a similar phenomena in the physical space found in \cite{kagan}.

We also investigate the controversial issue concerning the possible non-ergodic delocalized state of the disordered system. There is no definite viewpoint concerning this issue in the literature. To shed the light on this question in our CERN and RRG models, we compare their spectral properties in the clustered phase with the spectral properties of specially prepared reference model which has the same geometrical properties but has no pre-history. It turns out that although both models are in the delocalized phase in the main zone their spectral densities are very different. Hence the modes in the main zone strongly depends on initial conditions, thus indicating their non-ergodicity.

We conjecture that 3-cycles in the Fock space can be associated with resonant triples in the Hilbert space of many-body system considered in the transport phenomena \cite{triple}. Hence by tuning the chemical potential, $\mu$, of 3-cycles we can investigate the influence of resonant triples in the Hilbert space on the transport properties. Using the mapping of the Fock space onto the physical space (developed in \cite{kamenev,basko}) together with the localization transition (found in our work), we argue that clusters in the Fock space can be regarded as the degrees of freedom in the initial system.

The paper is organized as follows. The spectral properties of the modes in the main and non-perturbative zones are found numerically in Section 2. We also suggest a simple mean field qualitative explanation of the defragmentation phenomena at critical $\mu$. In Section 3 we discuss the interpretation of our numerical findings in terms of interacting many-body system in the physical space. In Section 4 we pay attention to the numerical evidence of the non-ergodic nature of the delocalized modes. Open questions are outlined in the Discussion.

\section{Level spacing distribution in perturbed CERN and RRG models}

\subsection{Numerical results and their interpretation}

To begin with, let us recall the critical behavior in CERNs \cite{crit2} which generalizes the one of the Strauss model \cite{strauss}, described in the mean-field approximation in \cite{burda,newman}. The CERN differs from the standard Erd\H{o}s-R\'enyi (ER) network by two additional requirements: i) the conservation of the vertex degree in all graph nodes, and ii) the condition that network is forced to increase the number of 3-cycles, $N_{\triangle}$.

The initial state of the constrained Erd\H{o}s-R\'enyi network is prepared by connecting any randomly taken pair of vertices with the probability $p$ (the double connections are excluded). After the initial pattern is prepared, one randomly chooses two arbitrary links, say, $(ij)$ (between vertices $i$ and $j$) and $(km)$ (between $k$ and $m$), and reconnect them, getting new links $(ik)$ and $(jm)$. Such a reconnection conserves the vertex degree \cite{maslov}. Now, one applies the standard Metropolis algorithm with the following rules: a) if under the reconnection the number of 3-cycles is increased, a move is accepted, b) if the number of 3-cycles is decreased by $\Delta N_{\triangle}$, or remains unchanged, a move is accepted with the probability $e^{-\mu \Delta N_{\triangle}}$. Then the Metropolis algorithm runs repeatedly for large set of randomly chosen pairs of links, until it converges. In \cite{reconnection} it was proven that the Metropolis algorithm converges to the true ground state $e^{\mu N_{\triangle}}$ in the equilibrium ensemble of random undirected constrained Erd\H{o}s-R\'enyi networks with particular values of vertex degrees in all nodes. Let us emphasize that the ground states of constrained and non-constrained Erd\H{o}s-R\'enyi networks are essentially different, which is reflected in non-ergodic behavior of CERNs typical for disordered systems with quenched disorder.

To explain the numeric results obtained for CERN and RRG, it is instructive to roll one step back and consider ensemble of unconstrained (i.e. conventional) Erd\H{o}s-R\'enyi graphs with non-conserved vertex degree. The Hamiltonian for such system reads
\be
H=-\mu N_{\Delta}, \qquad (\mu>0)
\label{eq:04}
\ee
which is the Hamiltonian of the Strauss model \cite{strauss,burda,newman}. It has been shown in
\cite{burda,newman} that if the system tends to form as many 3-cycles as possible, there is a phase
transition, driven by the chemical potential $\mu$, and at some $\mu_{cr}(p)$ the so-called Strauss phase consisting of a single full subgraph (clique) is formed for any initial random ER networks with fixed number of vertices, $N$, and fixed average number of links at the vertex, $pN$.

\begin{figure}[ht]
\centerline{\includegraphics[width=14cm]{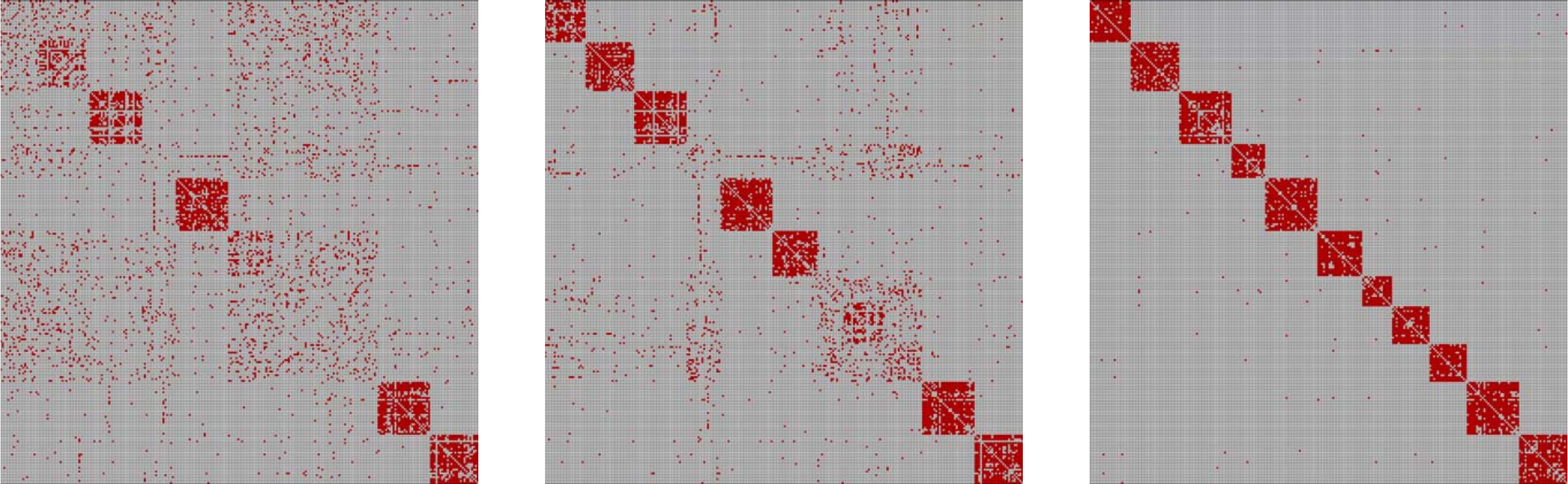}}
\caption{Few typical samples of intermediate stages of CERN evolution above the transition point.}
\label{fig:01}
\end{figure}

If the additional constraint of vertex degree conservation is imposed, such a model (CERN) becomes in many respects similar to the random regular graph. The phase transition which happens in CERN and RRG models is controlled by $\mu$, however networks decay into the maximally possible number, $[p^{-1}]$, of almost full subgraphs (cliques) \cite{crit2}, where $p$ is the vertex connection  probability at the preparation and $[...]$ denotes the "integer part". The corresponding adjacency matrices are shown in the \fig{fig:01}, where different phases of the ground states are clearly seen. The ground state of the quenched network involves $[p^{-1}]$ almost complete graphs corresponding to blocks (cliques) of the adjacency matrix $A$ with fluctuating sizes $N_i$ ($\sum_i N_i = N$) and the mean value of the vertices in the clique, $N_{cl}=\left<N_i\right>= N/[p^{-1}] \approx Np$. To visualize the kinetics, we enumerated vertices at the preparation condition in arbitrary order and run the Metropolis stochastic dynamics. After equilibrating the system, and when the cliques are formed, we re-enumerate vertices according to their belongings to cliques. Then we restore corresponding dynamic pathways back to the initial configuration.

Since we are interested in detailed description of the localization phenomena, the enveloping shape of the spectral density of adjacency matrix, and the level spacing distribution, become of primary importance. The spectral densities of ensembles of CERN and RRG behave similarly above $\mu_{cr}$ (see \cite{crit2} for details). Namely, when $\mu$ is increasing from zero, the Wigner semicircle gets gradually deformed, then it acquires the triangular shape of the main zone, accompanied by number of isolated eigenvalues, which eventually form the second zone. Isolated eigenvalues correspond to clusters \cite{newman1}, hence we can conjecture that the second zone is constituted of "non-perturbative" clusters with \emph{inter-cluster} interactions, while the spectral density in the main zone reflects the "perturbative" excitations \emph{inside} each cluster.

The investigation of the level spacing distribution for CERN and RRG provides the identification of the Anderson localization for the one-particle dynamics in the Fock space. It is known \cite{mehta} that in the delocalized regime, the distribution is given by the Wigner surmise, while in the localized phase it becomes Poissonian:
\be
\left\{
\begin{array}{ll}
P_{deloc}(s)\approx e^{-s^2/\sigma^2} & \mbox{below mobility edge, $\lambda_m$ (GOE)}
\medskip \\ P_{loc}(s) \approx e^{-s/\delta} & \mbox{above mobility edge, $\lambda_m$}
\end{array} \right.
\label{eq:05}
\ee
where $\sigma$, $\delta$ are some positive constants and GOE is the abbreviation of the Gaussian Orthogonal Ensemble, $s=\frac{\lambda_i -\lambda_{i+1}}{\Delta}$.

\begin{figure}[ht]
\centerline{\includegraphics[width=14cm]{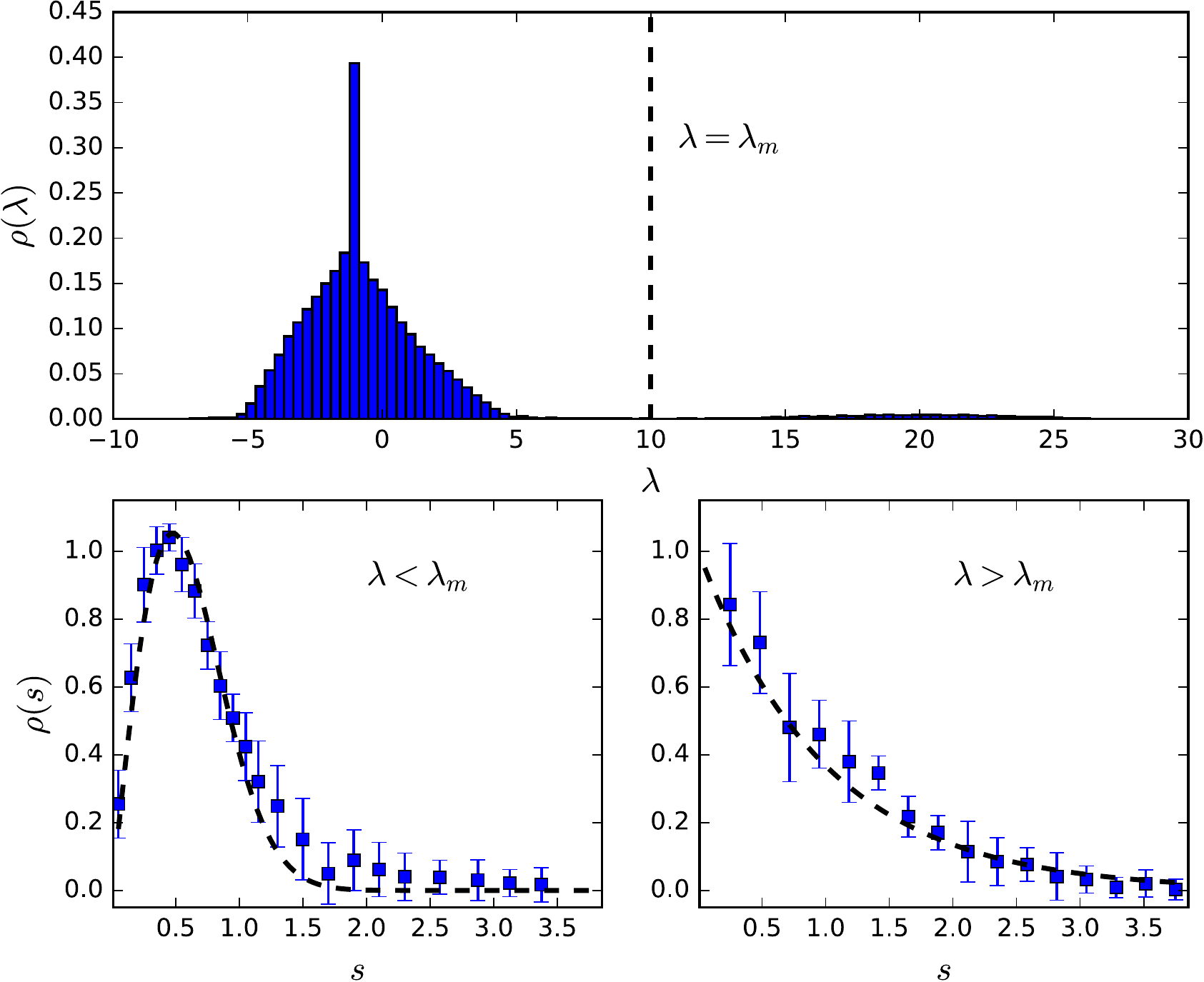}}
\caption{Spectral density and level spacing distribution for CERN,N=250, p=0.08,
averaging over 2500 realizations.}
\label{fig:02}

\end{figure}

\begin{figure}[ht]
\centerline{\includegraphics[width=14cm]{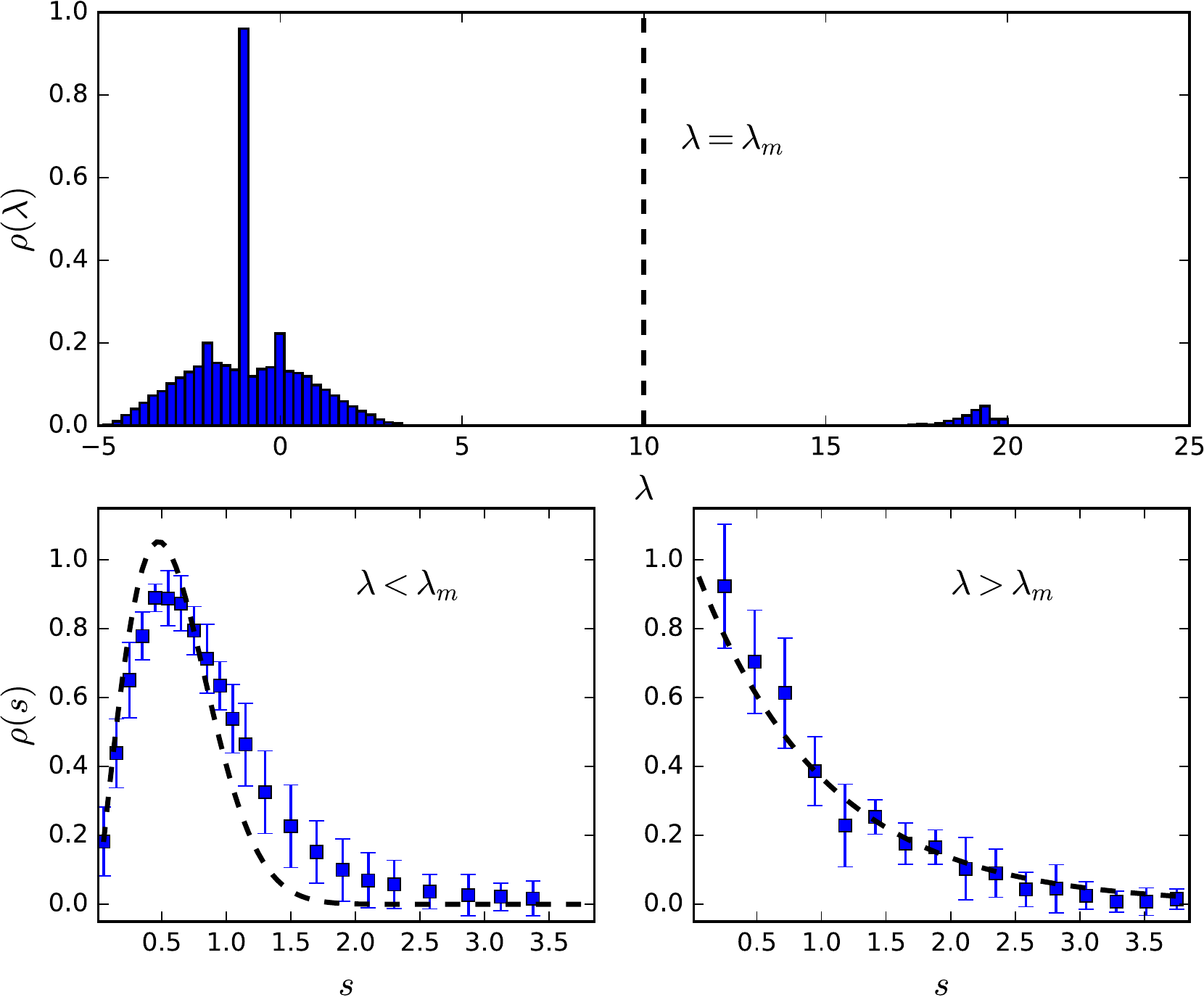}}
\caption{Spectral density and level spacing distribution for RRG,
N=250, d=20, averaging over 2500 realizations.}
\label{fig:03}
\end{figure}

In our work we have studied how the defragmentation of the CERN and RRG into clusters behind the phase transition in $\mu$, is accompanied by changes in the spectral densities and in the level spacing distributions. The results of our numerical simulations are shown in the \fig{fig:02} for CERN, and in the \fig{fig:03} for RRG. It is seen for both, CERN and RRG, that the level spacing distribution in the central zone follows the Wigner surmise, while it becomes Poissonian for the eigenvalues in the second zone. These distributions are separated by the well defined mobility edge, $\lambda_m$, which in our case lies between two zones. The second zone corresponds to the localized phase and can be naively interpreted in physical terms as the insulator. Fine structure of delocalized states in the main zone and its meaning will be discussed later.


Since the mobility edge separates the localization-delocalization regimes in the Fock space, we can ask ourselves how our findings can be interpreted for the initial many-body system. Following arguments of \cite{kamenev}, it seems naturally to identify the Fock space states in the localized part of the spectrum (the second zone) with the localized states in the initial many-body interacting model. Since we know that the localized state in the Fock space is the cluster formed in CERN or RRG above the transition point $\mu_{cr}$, we come to the straightforward identification of the network cluster with the \emph{quasiparticle} in the underlying many-body system in the coordinate space for which CERN or RRG parametrize the Fock spaces.

We argue that in the localized regime in the CERN and RRG, the eigenstates in the Fock space are identified with clusters. According to the logic of the many-body localization, this means that beyond the mobility edge clusters form the proper basis in the initial many-body system at $\mu>\mu_{crit}$ and can be considered as the corresponding degrees of freedom (quasiparticles). However, in the delocalized regime below the mobility edge (note that the system is still kept at $\mu>\mu_{crit}$), the perturbative excitation in the main zone of the spectral density should be considered as the collective state of large number of interacting clusters.

Below the phase transition (at $\mu<\mu_{crit}$) such a "cluster basis" does not exist since we have no stable clusters in the Fock space representation in this regime. It is worth noting that similar situation is familiar for systems which enjoy different spectra of stable particles at weak and strong coupling regimes. At strong coupling, the natural basis involves the soliton-like states, and perturbative modes have to be considered as coherent state of solitonic states. Contrary, at weak couplings, the natural basis involves perturbative modes and solitons are considered as made from large number of perturbative modes.

The existence of the mobility edge has straightforward explanation in this model picture. In the delocalized region of the spectrum we have only perturbative modes which can be considered as excitations of solitonic state. According to \cite{crit2}, the interaction between solitons-clusters has dramatic influence on the spectral density: instead of separated peaks which constitute the spectrum of almost full (or dual sparse) networks, we have the continuum set of modes which has very peculiar triangular enveloping shape. One could say that perturbative modes get collectivized among the solitons-clusters.

\section{Interpretation: localization without disorder in the Fock space}

Recall that in our model we have no on-site (diagonal) disorder in the Fock space, hence it is crucial to understand better the localization mechanism. The proper framework is provided by the scenario of localization without disorder in translationally invariant systems first suggested in \cite{kagan} and reconsidered recently in \cite{huv1,huv2} from the perspective of many-body localization. In the original work \cite{kagan} the resonant spots located in the physical space, produce an effective structural disorder and play the key role for the localization phenomena. This mechanism crucially differs from localization in systems with the diagonal disorder, where  possible resonant spots are placed in positions dictated by the disorder distribution. For  translationally invariant systems, positions of rare spots in the physical space are not fixed and they are potential carriers in the transport phenomena. Inside resonant spots the states are delocalized. In models considered in \cite{huv1,huv2}, the resonant spots are induced by the temperature and a bit loosely one could claim that the "randomness" is induced by the initial state.

The localization phenomena we have found in CERNs and RRGs has a lot in common with this scenario. The translationally invariant system we are dealing with, and which enjoys the one-particle localization, lives in the Fock space like in the situation with the on-side disorder. Again, we can treat the one-particle localization in the Fock space as the manifestation of the many-body localized (MBL) states in the physical space. In our situation the resonant spots are induced not by the temperature, but by the chemical potential $\mu$ of 3-cycles. The clusterization occurs at $\mu \ge \mu_{crit}$ and the states involved into some resonant spot form a cluster. The spectrum inside the resonant spot corresponds to delocalized states -- exactly as in \cite{kagan, huv1,huv2}. The dependence on the initial state occurs naturally in our model.

Let us emphasize that in CERNs and RRGs we have two sources of the disorder. On one hand, due to the network ramdomness at the preparation, we have the non-diagonal disorder for any configuration of 3-cycles and any $\mu$. On the other hand, due to the clusterization happen at $\mu\ge \mu_{crit}$, we induce a structural disorder. One could ask whether the non-diagonal disorder can produce the one-particle localization in the Fock space. The answer is negative: it was shown in \cite{crit1} that there are no clusters at $\mu\le \mu_{crit}$ and all states in the system are delocalized. The clusterization occurs for any initial state only for $\mu \ge \mu_{crit}$. Hence, the off-diagonal disorder is not entirely responsible for the clusterization. Let us note that the localization phenomena found in CERNs essentially differs with the localization forced by the diagonal disorder in conventional Erd\H{o}s-R\'enyi networks \cite{slanina}.

Since the structural disorder in the Fock space is induced by the chemical potential of 3-cycles, $\mu$, it is worth understanding its physical meaning. To clarify the situation, let us remind the notion of resonant triples discussed in \cite{triple}. The resonant triple is defined as the triple of states in the Hilbert space of some quantum interacting many-body system obeying mutual resonant conditions. The resonant condition means that the difference between energies of three states in the absence of interaction is small compared to the effects of interaction. There is also the semiclassical analog of the resonant triple which corresponds to the peculiar behavior of the separatrix width and the Arnold diffusion \cite{triple} (see also \cite{huse2009}). The resonant triples yield the thermal chaotic spots \cite{triple} which amount to the transport in the system.

In the network model of the Fock space a link corresponds to a resonant pair, while a 3-cycle is a resonant triple. The estimation of the number of the resonant triples in the Hilbert space of the real interacting many body system is a very complicated issue \cite{triple, burin} and there is no well defined answer yet. Our study shows that the number of resonant triples is crucial for properties of the many-body system, and especially it influences localization and transport phenomena. As the number of resonant triples exceeds some critical threshold, the Fock space of the interacting many-body system gets clusterized. Note the although yet we have investigated the impact of the 3-cycles only, we expect that higher cycles (resonant quarters, etc) play the similar role, affecting the structural disorder above the critical value of the corresponding chemical potential.

\section{Triangle-shaped spectral density as a signature of non-ergodic delocalized phase}

Here we discuss the question whether the main zone of delocalized states is in ergodic or non-ergodic state. The signature of the non-ergodic behavior of the spectral density in the delocalized phase formulated in the studies related to this issue \cite{biroli,krav}, is as follows. The spectral density does not share the Wigner semicircle law (as in the central zone of CERN and RRG), however the level spacing distribution still is of the Wigner-Dyson type.

\begin{figure}[ht]
\centerline{\includegraphics[width=14cm]{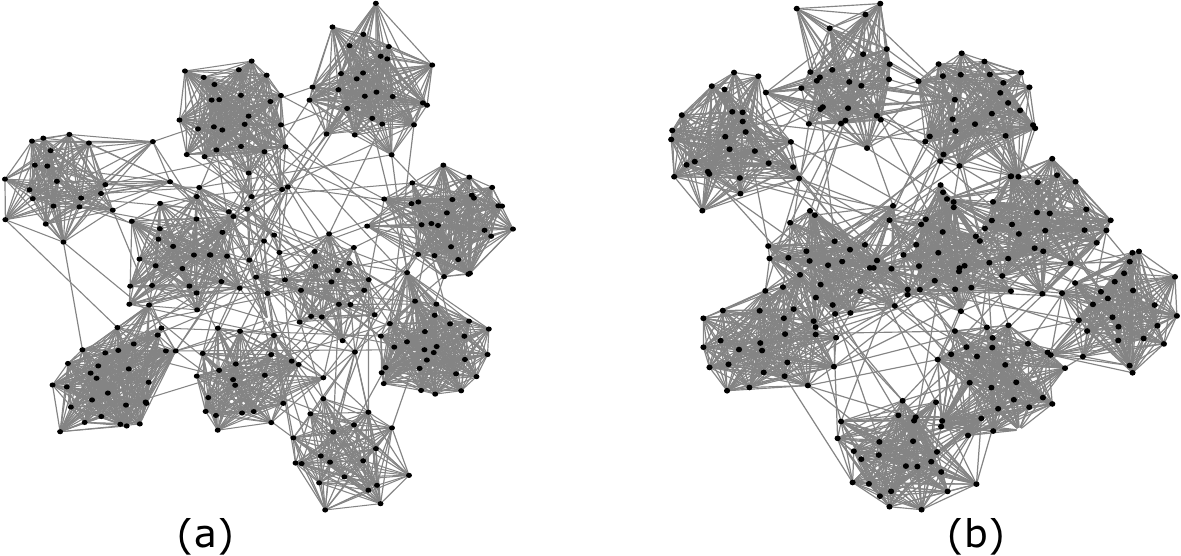}}
\caption{Visualizations of: (a) evolutionary prepared CERN, and (b) "memory-independent" CERN prepared \emph{ad hoc}. Both networks have $M=10$ clusters and have effective connection probabilities $p_{in}=0.8$ in clusters and $p_{out}=0.004$ between clusters. Despite visual resemblance, the spectra of both networks are essentially different.}
\label{fig:04}
\end{figure}

To address this issue, we considered the spectral density of the network consisting of clusters prepared \emph{ad hoc} with exactly the same connectivity parameters as in the CERN, results of which are depicted in \fig{fig:01}. Namely, after the evolution of the CERN, maximizing the number of 3-cycles, the particular final network (fCERN) looks (see \fig{fig:04}a) as a collection of $M=10$ weakly connected dense clusters. Thus, for the fCERN we can define the effective inter-- and intra-- cluster connectivities ($p_{in}\approx 0.8$ and $p_{out}\approx 0.004$). Now we can mimic fCERN by connecting nodes \emph{ad hoc} with different probabilities: \emph{inside} each cluster with the probability $p_{in}=0.8$, and between \emph{different} clusters with the probability $p_{out}=0.004$. Such an "instantly created" network is depicted in \fig{fig:04}b. Such a network does not have any pre-history, it knows nothing about the evolution, it has no dependence on the chemical potential of 3-cycles, and it has no any constraints.

\begin{figure}[ht]
\centerline{\includegraphics[width=14cm]{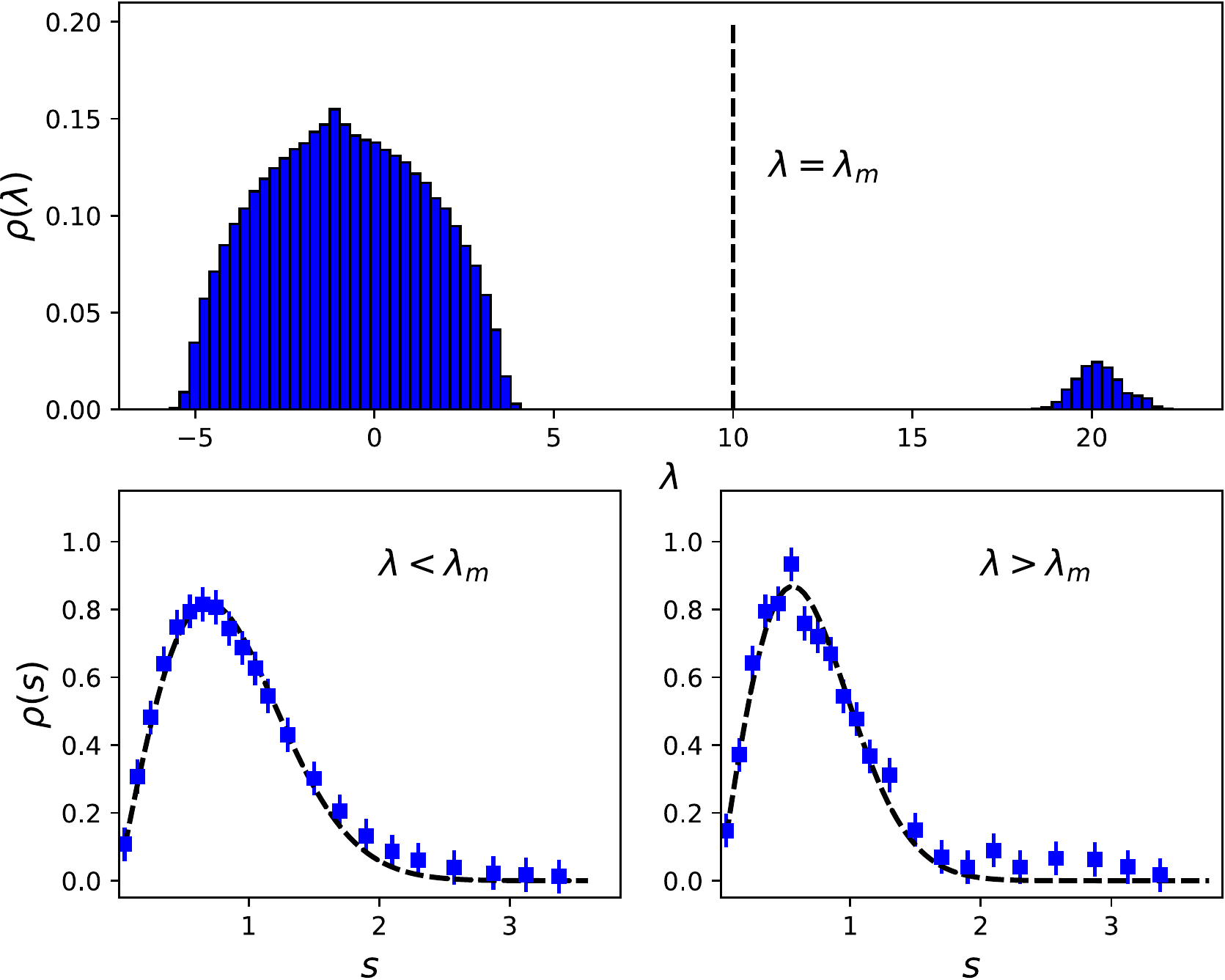}}
\caption{Spectral density and level spacing distribution for the "memory-independent" \emph{ad hoc} prepared network with $M=10$ clusters and $p_{in}=0.8$, $p_{out}=0.004$. There is no mobility edge in the system: all eigenmodes are completely delocalized in both zones.}
\label{fig:05}
\end{figure}

The instantly created network is a perfect reference system without additional constraints and we could expect that corresponding modes are completely delocalized. The spectrum of this model is presented in \fig{fig:05} and we see that indeed all eigenvalues are delocalized, despite there is no any visual difference between the network which is the final state of the evolution of CERN (\fig{fig:04}a), and the instantly created "static" random network (\fig{fig:04}b).

Now we are interested whether the states of CERN after the evolution are ergodic or not. Let us compare the spectral density of the \emph{memory-independent} instantly created random network, shown in \fig{fig:05} with the one of the \emph{memory-dependent} final state of CERN, depicted in \fig{fig:02}. We clearly see that the network with \emph{ad hoc} prepared clusters does not have the triangle enveloping shape of the spectral density, however shares the Wigner surmise for the level spacing distributions in both zones. Thus we can expect that the states in the CERN and RRG are non-ergodic being essentially "memory-dependent", and the triangular shape of the central zone of the spectral density presumably is its signature since it emerges only at the final stage of the evolution maximizing 3-cycles.

The emergence of the nonergodicity of the delocalized phase has been advocated in \cite{biroli,krav,delu, biroli2,kkca} and is the subject of intensive (and sometimes controversial) discussions. In \cite{tikhonov} it was conjectured that in the large-$N$ limit such a non-ergodic behavior disappears being a finite-size effect. However, it was argued in \cite{ioffe} that the emergence of the non-ergodic phase is supplemented by the one-step replica symmetry breaking (RSB) and the transition between ergodic and non-ergodic phases becomes even more sharp in the large-$N$ limit, being the real phase transition separating the replica symmetric phase and the phase with broken replica symmetry. The block-diagonal structure of the CERN adjacency matrix shown in the last panel of \fig{fig:01} implies that we have a kind of a one-step replica symmetry breaking for the whole network. This allows us to claim that clusterization in the CERN at $\mu_{cr}$ is accompanied by the occurrence of non-ergodic "memory-dependent" behavior of the entire system. It is worth noting that the very fact of presence of the vertex degree conservation has much stronger impact on statistics of the rewired network than just the inhomogeneity of vertex degrees in different network nodes. That is why the statistical behavior of CERNs and RRGs is almost identical up to some minor details.

The strong memory dependence of CERNs observed in simulations, allows to conjecture that CERNs and RRGs have some hidden nonlocal conservation laws, which is a common feature of the many-body localization pattern \cite{ives}. There are natural candidates for hidden integrals of motion in our study. Indeed we impose the condition of the degree conservation in all vertices, hence we have $N$ conserved integrals of motion. They are expressed non-locally in terms of natural degrees of freedom -- links of the network. The number of conserved quantities is essentially smaller than the total number of degrees of freedom in the network, hence we do not have a complete integrability. It would be interesting to investigate the situation where the degree conservation constraints are imposed on the part of the vertices only. Certainly, this point deserves further detailed study.

\section{Conclusion}

In this paper we have analyzed the localization properties of perturbed  constrained Erd\H{o}s-R\'enyi networks and random regular graphs. In our model in the initial network we have neither diagonal disorder, nor the off-diagonal intrinsic disorder, which could cause localization. The localized phase emerges due to the structural disorder spontaneously induced above some critical chemical potential $\mu$ for the 3-cycles. This is the realization of the scenario "localization without disorder" in the translationally invariant systems \cite{kagan}. Since we interpret our network as the model for the Fock space of the interacting many-body system, it is crucial to identify the meaning of the 3-cycles in the Fock space. We suggest that they represent the resonant triples \cite{triple}, and our study indicates that the number of such resonant triples strongly influences the properties of the system.

We have shown numerically that above the critical value of the chemical potential, $\mu_{cr}$, the spectrum on the network gets splitted into zones of delocalized and localized modes separated by the mobility edge. The localized modes in the second zone correspond to clusters, while the delocalized modes of the main zone are identified with the perturbative excitations around the clusters and are collectivized via inter-cluster interactions. We present arguments based on direct visualization of adjacency matrix structure, favoring the conjecture that the modes above $\mu_{cr}$ are in the non-ergodic regime enjoying a kind of one-step replica symmetry breaking. We have focused in our paper mainly on the level distribution statistics. It would be useful to extend this analysis to the participation ratio, the entanglement entropy \cite{grover}, or other benchmarks of the localization transition. We could mention also the many-body counterpart of the Thouless energy \cite{abanin,mon}.

We believe that the mechanism of CERN and RRG defragmentation via of eigenvalue tunnelling \cite{crit2}, when the system is pushed to some "untypical" topological state with a large number of 3-cycles, is quite general. The eigenvalue tunnelling we are dealing with, has various incarnations is physics. For instance, formation of the "eigenvalue instantons" in many physical situations signals the creation of the stable solitons (see \cite{marino} for review). Our numerical study shows that in such a model of interacting clusters, the mobility edge can be formed, and the identification of the initial and effective degrees of freedom is very different at two sides of the mobility edge.

We focused only on "colorless" CERN and RRG where vertices do not carry any additional degrees of freedom. In \cite{crit1,crit3} the similar analysis for the colored network has been developed, where it was shown that some additional color-dependent critical phenomena can occur.

In the model considered here, there are two types of natural degrees of freedom: clusters and elementary links. We have argued that at strong couplings the fundamental degrees of freedom, network cliques (clusters), can be identified with the soliton-like quasiparticles in the initial interacting many-body problem. It is natural to conjecture a kind of a weak-strong particle-soliton duality behind the many-body localization. Such a picture is consistent with the genus-one Riemann surface for the two-zonal spectral density. The approach dealing with the modular structure behind the Anderson transition developed in \cite{kamenev2} seems to be relevant for the identification of the modular structure in the many-body localization in real space and corresponding Anderson localization in the Fock space. Another incarnation of the modular invariance in localization phenomena for ensemble of sparse matrices, is discussed in \cite{krapiv}.

We are grateful to M. Hovhannisyan, A. Kamenev, V. Kravtsov, V. Pasquier, N. Prokof'ev and M. Tamm for many useful discussions.  The work of A.G. was performed at the Institute for Information Transmission Problems with the financial support of the Russian Science Foundation (Grant No.14-50-00150). S.N. acknowledges the support of the IRSES DIONICOS grant and of the RFBR grant No. 16-02-00252. O.V. thanks Basis Foundation Fellowship for support.The
work of V.A. was supported within frameworks of the state task for ICP RAS 0082-2014-0001
(state registration AAAA-A17-117040610310-6)

\end{document}